\documentclass[conference]{IEEEtran}
\IEEEoverridecommandlockouts
\usepackage{cite}
\usepackage{amsmath,amssymb,amsfonts}
\usepackage{graphicx}
\usepackage{xcolor}
\usepackage{url}
\usepackage{booktabs}
\usepackage{subcaption}
\def\BibTeX{{\rm B\kern-.05em{\sc i\kern-.025em b}\kern-.08em
    T\kern-.1667em\lower.7ex\hbox{E}\kern-.125emX}}

\begin{document}

\title{On the Geometric Limits of Transformer Defenses against Obfuscation Attacks: Latent Embedding Collapse \& Performance–Robustness Gap}

\author{\IEEEauthorblockN{Becky Mashaido and Tapadhir Das}
\IEEEauthorblockA{Department of Computer Science, University of the Pacific, Stockton, USA}
Email: b\_mashaido@u.pacific.edu, tdas@pacific.edu}

\maketitle
\begin{abstract}
Prompt-injection attacks pose significant risks to language model safety, yet existing defenses are typically evaluated using classification performance. We show that high detection performance does not imply representational robustness. Specifically, multi-operator obfuscated prompts (combining homoglyphs, zero-width characters, and punctuation or emoji noise) can partially collapse onto the embedding manifold of clean prompts, a phenomenon we term $latent$ $embedding$ $ collapse$. Results indicate that across multiple BERT-family encoders with varying depth and capacity, detectors achieve near-perfect classification performance, yet the minimal clean–obfuscated margin $\delta = 1.02$, indicating near-overlap of obfuscated and clean embeddings. Obfuscated embeddings further exhibit elevated intra-class variance ($3.33 \pm 6.23$), indicating severe latent-space instability despite high performance. These results reveal a substantial $performance–robustness$ $gap$, demonstrating that standard evaluation metrics fail to capture latent embedding collapse and underlying geometric fragility. Our findings show that increasing model capacity does not eliminate latent embedding collapse, motivating geometry-aware robustness analysis as a necessary complement to performance-based evaluation for prompt-injection defenses.
\end{abstract}


\section{Introduction}
\begin{figure}[t]
\centering
\includegraphics[width=\columnwidth]{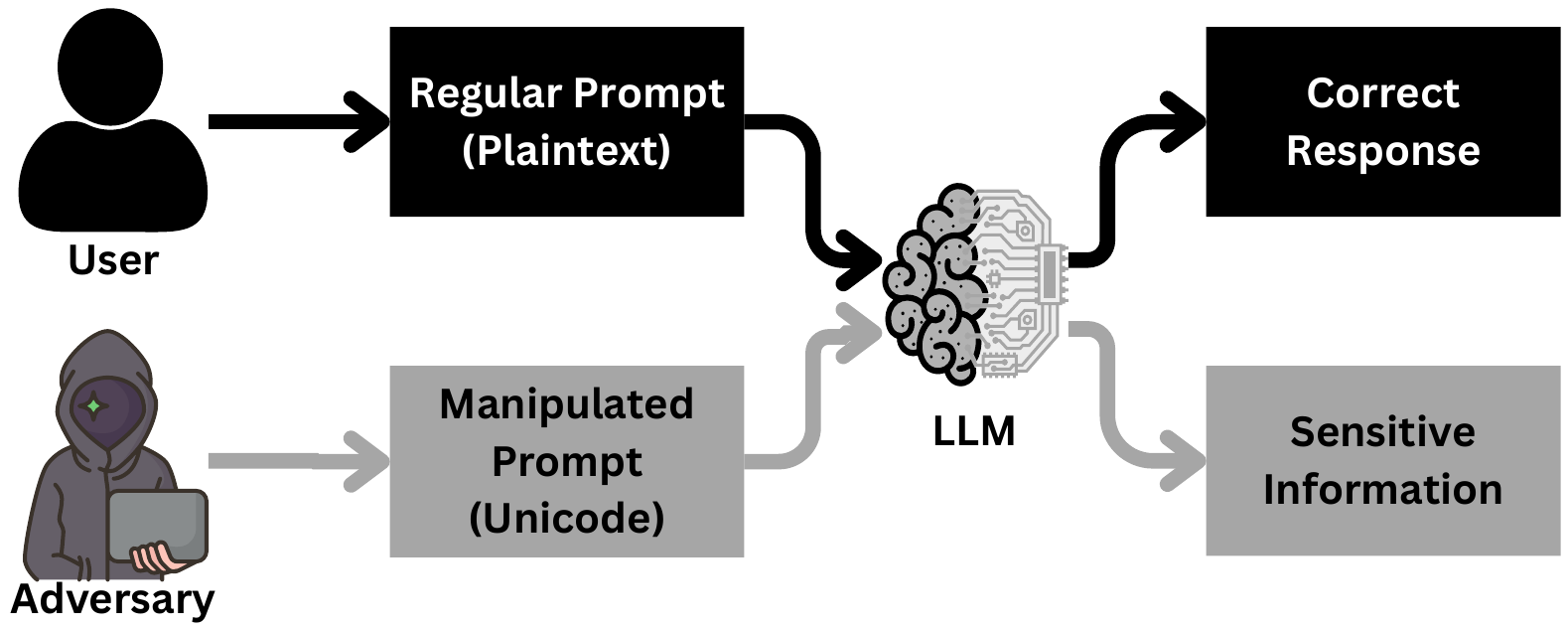}
\caption{Illustration of Prompt Injection Attack on LLMs}
\label{fig:prompt}
\end{figure}

Large language models (LLMs) have become a ubiquitous part of our daily personal \& professional lives. These are power neural network models that are trained on a wide variety of text and code to comprehend, reproduce, and handle human language \cite{kasneci2023chatgpt}. Due to their widespread usage, the market size of LLMs are expected to hit \$35.43 billion by 2030 \cite{grandviewresearchLargeLanguage}. However, despite their extended use, LLMs can be compromised by cyber attacks such as prompt injections \cite{liu2023prompt}. Prompt-injection attacks pose significant risks to the safety of LLMs by embedding malicious instructions within seemingly benign user inputs. Such attacks can override system directives, manipulate model behavior, or exfiltrate sensitive information without explicit policy violations \cite{shi2024optimization}. An illustration of prompt injection can be seen in Figure \ref{fig:prompt}. 

As a result, a growing body of work proposes prompt-injection detection mechanisms based on supervised classification, typically evaluated using accuracy, precision, recall, or F1 score~\cite{hung2024attention, chen2025defending}. While these metrics capture observable misclassification behavior, they provide limited insights into the robustness of the underlying representations learned by transformers. In particular, a detector may achieve near-perfect performance while relying on fragile latent representations that can be exploited by adaptive adversaries. This raises a fundamental question: \emph{does high detection performance imply representational robustness against compositional prompt obfuscation?}

In this paper, we have investigated this research question and have discovered that the answer is \textbf{No}. For this, we demonstrate that multi-operator obfuscated prompts (constructed through combinations of Unicode homoglyphs, zero-width characters, fragmented instructions and punctuation or emoji noise) can intrude into regions of the latent space occupied by clean prompts even when classification performance remains near-perfect. We refer to this phenomenon as \textbf{latent embedding collapse}, characterized by a partial manifold collapse in which obfuscated prompts intrude into regions of the latent space occupied by clean prompts without inducing misclassifications. To the best of our knowledge, this is the first research that is studying embedding-space or geometric analyses in LLM adversarial defense. This makes our focus on latent embedding collapse a novel perspective. The main contributions of this work are threefold:
\begin{enumerate}
    \item Formalizing the concept of \emph{latent embedding collapse} and connecting it to partial manifold collapse in transformer embedding spaces.
    \item Introducing geometric metrics to quantify representational robustness, providing a complementary evaluation to conventional performance metrics.
    \item Demonstrating empirically that increasing model depth or capacity does not resolve latent embedding collapse, indicating that the \textbf{performance-robustness gap} is an inherent phenomenon rather than an artifact of limited model size.
\end{enumerate}
The rest of the paper is structured as follows: Section~\ref{sec:related} provides our literature review; Section ~\ref{sec:sys} introduces the system model; Section~\ref{sec:methodology} details the methodology of the research; Section~\ref{sec:evaluation} presents evaluation results for our project; and  Section~\ref{sec:conclusion} provides concluding remarks and future work.

\section{Related Work} \label{sec:related}
Prompt-injection attacks in LLMs have attracted increasing attention due to their potential to subtly alter model outputs \cite{rahman2025fine} \cite{chang2025chatinject} \cite{greshake2023not} \cite{lee2024prompt}. The work in \cite{rahman2025fine} explored the security vulnerabilities in relation to prompt injection attacks for LLMs. Researchers in \cite{chang2025chatinject} introduced ChatInject, an attack that formatted malicious payloads to mimic native chat templates in LLMs. The work in \cite{greshake2023not} demonstrated that LLMs blur the lines between data and instructions during prompt injection attacks. Authors in \cite{lee2024prompt} introduced Prompt Infection, where malicious prompts self-replicate across interconnected LLM agents. 

Prior defenses largely adopt a supervised classification approach, detecting malicious prompts using accuracy, precision, recall, or F1 score as evaluation metrics \cite{lin2025uniguardian} \cite{hung2024attention} \cite{chen2025defending} \cite{piet2024jatmo}.In \cite{lin2025uniguardian}, authors proposed UniGuardian, the first defense designed to detect prompt injection, backdoor attacks, and adversarial attacks in LLMs. The work in \cite{hung2024attention} investigated the underlying mechanisms of prompt injection attacks by analyzing the attention patterns within LLMs. Researchers in \cite{chen2025defending} proposed DefensiveToken, a defense with prompt injection robustness comparable to training-time alternatives. Authors in \cite{piet2024jatmo} introduced Jatmo, a defense mechanism for generating task-specific models resilient to prompt injection attacks. Through these works, we observe that performance metrics demonstrate strong detection performance. However, a limitation of these works is that they provide limited insight into the geometry of the learned latent space and the robustness of underlying representations \cite{goyal2022survey}. 

Several surveys and critiques highlight the limitations of performance-centric evaluation in this domain \cite{wang2025review} \cite{jia2025critique}. Even recently proposed defenses, including data filtering \cite{wang2025datafilter} and representation-based methods \cite{liu2025drip}, remain primarily evaluated on observable classification outcomes, rather than examining latent embedding structure. Observations on embedding-space or geometric analyses are rare in adversarial defense, making our focus on latent embedding collapse a novel perspective. This gap motivates our approach: rather than focusing solely on misclassification rates, we analyze the \emph{latent embedding collapse}, a phenomenon in which obfuscated prompts partially intrude into regions of the latent space occupied by clean prompts without inducing misclassification. By emphasizing geometric properties of transformer representations, our method complements prior performance-based defenses and provides deeper insight into model robustness, revealing vulnerabilities that conventional metrics overlook.

\section{System Model} \label{sec:sys}
In this section, we are providing the mathematical background for latent embedding collapse. Let $\phi(x) \in \mathbb{R}^d$ denote the final-layer embedding of a prompt $x$ produced by a transformer encoder. For each prompt class $c \in \{\mathrm{clean}, \mathrm{prefix}, \mathrm{suffix}, \mathrm{obfuscated}\}$, we define the corresponding embedding manifold $M_c = \{\phi(x) \mid x \sim P_c\}$. Latent embedding collapse occurs when obfuscated prompts occupy regions of the embedding space arbitrarily close to those of clean prompts, resulting in a loss of effective geometric separation between manifolds.

We characterize this collapse through two complementary geometric conditions. First, the \emph{clean-obfuscated margin}
\begin{equation}
\delta =
\min_{x_c \in M_{\mathrm{clean}},\, x_o \in M_{\mathrm{obf}}}
\|\phi(x_c) - \phi(x_o)\|_2
\end{equation}
quantifies the minimal distance between clean and obfuscated embeddings. A small $\delta$ indicates that some obfuscated prompts lie extremely close to clean prompts in the latent space. Second, the \emph{obfuscated intra-class variance}
\begin{equation}
\sigma^2_{\mathrm{obf}} =
\frac{1}{|M_{\mathrm{obf}}|^2}
\sum_{x_i, x_j \in M_{\mathrm{obf}}}
\|\phi(x_i) - \phi(x_j)\|_2^2
\end{equation}
captures the degree of dispersion among obfuscated embeddings, reflecting latent instability and spread across both benign and adversarial regions.

We formalize this discrepancy as an \textbf{performance-robustness gap}, wherein high detection performance coexists with severe embedding-space vulnerability. By reframing prompt-injection detection as a problem of representation geometry rather than classification alone, this work highlights a fundamental limitation of existing defenses and motivates geometry-aware robustness evaluation as a necessary complement to performance-based metrics.

\section{Methodology} \label{sec:methodology}
\begin{figure}[t]
\centering
\includegraphics[width=\columnwidth]{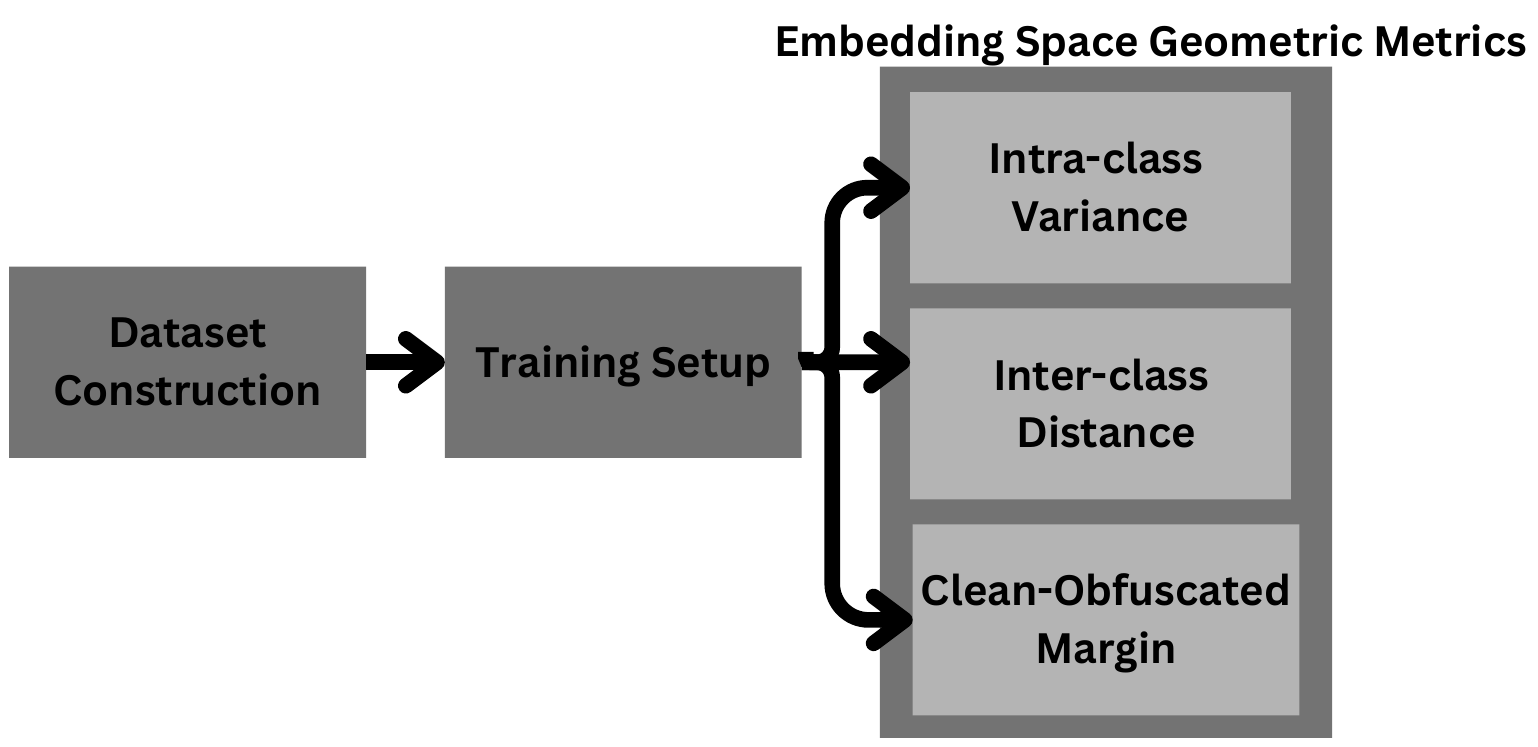}
\caption{Proposed methodology of understanding obfuscated prompt interaction with clean embeddings}
\label{fig:methods}
\end{figure}

In this section, we describe our methodology for this research. Our goal is to understand how obfuscated prompts interact with clean embeddings, revealing phenomena such as latent embedding collapse and partial manifold collapse. Our proposed methodology is provided in Figure \ref{fig:methods}.

\subsection{Dataset Construction}
We construct a dataset of 10,000 examples per label type across four classes: \textit{clean}, \textit{prefix}, \textit{suffix}, and \textit{obfuscated}. Clean sentences are generated from natural-language templates using nouns, verbs, adjectives, adverbs, and contextual extras. Prefix and suffix classes are created by prepending or appending partial or full instructional fragments, including both complete and fragmented instructions, to clean sentences. The obfuscated class is designed to probe \textit{latent embedding collapse} by applying multi-layer transformations to base sentences. These transformations include:
\begin{itemize}
    \item Unicode homoglyph substitutions, e.g., Cyrillic replacements for Latin characters.
    \item Zero-width character insertions and emoji/punctuation noise.
    \item Embedded partial or fragmented instructional prefixes and suffixes.
    \item Stochastic composition of all transformations, ensuring semi-ambiguous or hidden signals.
\end{itemize}
Formally, for each base sentence $x$, the obfuscated function $\mathcal{O}(x)$ applies the above transformations probabilistically, creating a diverse manifold $M_{\mathrm{obf}}$ that partially intrudes into the clean manifold $M_{\mathrm{clean}}$.

\subsection{Training Setup}
We fine-tune three transformer models: DistilBERT, BERTBase, and BERTMedium for 4-way classification. Key hyperparameters:
\begin{itemize}
    \item Dropout in classifier head: 0.1–0.2  
    \item Weight decay: 0.01  
    \item Learning rate: $3\times10^{-5}$  
    \item Epochs: 25 with early stopping (patience 4)  
    \item Batch size: 16 per device  
\end{itemize}

We evaluate using accuracy, precision, recall, and macro F1. The three-model performance summary is shown in Table~\ref{tab:cross_model_results}.

\subsection{Embedding-Space Geometry Metrics}
Let $\phi(x) \in \mathbb{R}^{d}$ denote the final-layer embedding of prompt $x$. For each class $c \in \{\text{clean}, \text{prefix}, \text{suffix}, \text{obfuscated}\}$, define the embedding manifold
\[
M_c = \{\phi(x) \mid x \sim P_c\}.
\]

These metrics are designed to quantify the geometric structure of latent embeddings, exposing phenomena that are invisible to standard performance metrics, such as latent embedding collapse and partial manifold collapse.

\subsubsection{Intra-Class Variance}
\begin{equation}
\sigma^2_{\mathrm{intra}}(c) = 
\frac{1}{|M_c|^2} \sum_{x_i, x_j \in M_c} \|\phi(x_i) - \phi(x_j)\|_2^2
\end{equation}

This measures how tightly the embeddings of a given class cluster in latent space. A small $\sigma^2_{\mathrm{intra}}$ indicates a compact manifold, suggesting the model consistently represents the class. Conversely, high intra-class variance may signal sensitivity to noise or obfuscation.

\subsubsection{Inter-Class Distance}
\begin{equation}
D(c_1, c_2) = 
\frac{1}{|M_{c_1}||M_{c_2}|} 
\sum_{x \in M_{c_1}} \sum_{y \in M_{c_2}} \|\phi(x) - \phi(y)\|_2
\end{equation}

The inter-class distance captures the separation between two class manifolds. Larger distances indicate well-separated classes in latent space, which generally improve robustness. Critically, when obfuscated prompts intrude into the clean manifold, $D(\text{clean}, \text{obfuscated})$ decreases, signaling latent embedding collapse.

\subsubsection{Clean-Obfuscated Margin}
\begin{equation}
\delta = 
\min_{\substack{x_c \in M_{\mathrm{clean}} \\ x_o \in M_{\mathrm{obf}}}} 
\|\phi(x_c) - \phi(x_o)\|_2
\end{equation}

The minimum distance between clean and obfuscated embeddings highlights the most extreme cases of intrusion. A small $\delta$ reveals partial manifold collapse, where obfuscated prompts lie in regions occupied by clean prompts without triggering misclassification. This phenomenon is central to our claim: \textbf{performance alone does not capture subtle vulnerabilities in latent space}. These metrics collectively allow us to quantify the geometry of latent embeddings, going beyond performance-centric evaluations. By tracking intra-class variance, inter-class distances, and the clean–obfuscated margin, we can detect hidden overlap between classes (latent embedding collapse). We can also identify potential risks for LLMs to adversarial robustness, through which we can guide architecture or training modifications to improve manifold separation. This geometry-based perspective exposes vulnerabilities that traditional metrics miss and motivates the visualizations in Section~\ref{sec:evaluation} showing obfuscated prompts partially intruding into clean embedding regions.

\section{Results} \label{sec:evaluation}
We now analyze classification performance and latent-space geometry across our three transformer architectures (DistilBERT, BERTBase and BERTMedium) to evaluate whether high detection performance corresponds to representational robustness under multi-operator prompt obfuscation. 

\begin{table}[t]
\centering
\caption{Classification performance across transformer architectures.}
\label{tab:cross_model_results}
\begin{tabular}{lcccc}
\toprule
Model & Accuracy & Precision & Recall & F1 \\
\midrule
DistilBERT  & 0.993 & 0.993 & 0.993 & 0.993 \\
BERTBase    & 0.994 & 0.994 & 0.994 & 0.994 \\
BERTMedium  & 0.993 & 0.993 & 0.993 & 0.993 \\
\bottomrule
\end{tabular}
\end{table}

\begin{table}[t]
\centering
\caption{Pairwise Euclidean distances (mean $\pm$ std) between final-layer
DistilBERT embeddings. Clean-obfuscated margin $\delta$ is shown below.}
\label{tab:distances}
\resizebox{\columnwidth}{!}{%
\begin{tabular}{lcccc}
\toprule
 & Clean & Suffix & Prefix & Obfuscated \\
\midrule
Clean       & 1.71 $\pm$ 0.84 & 25.65 $\pm$ 0.23 & 25.79 $\pm$ 0.17 & 24.34 $\pm$ 4.51 \\
Suffix      & 25.65 $\pm$ 0.23 & 1.21 $\pm$ 0.40 & 23.93 $\pm$ 0.18 & 25.05 $\pm$ 0.18 \\
Prefix      & 25.79 $\pm$ 0.17 & 23.93 $\pm$ 0.18 & 1.51 $\pm$ 0.60 & 25.15 $\pm$ 0.19 \\
Obfuscated  & 24.34 $\pm$ 4.51 & 25.05 $\pm$ 0.18 & 25.15 $\pm$ 0.19 & 3.33 $\pm$ 6.23 \\
\midrule
\multicolumn{5}{c}{Clean-Obfuscated Margin $\delta = 1.02$} \\
\bottomrule
\end{tabular}}
\end{table}

First, we analyze the operating performance of each of the 3 observed models, as shown in Table \ref{tab:cross_model_results}. We observe that all 3 models achieve near-identical classification performance, with accuracy, precision, recall, and macro F1 consistently around 99\%. This indicates that the models can efficiently detect clean, suffix, prefix, and obfuscated prompts. From there, we perform a deeper analyses of the final-layer embeddings for one of the models (DistlBERT) as demonstrated in Table ~\ref{tab:distances}, which summarizes pairwise Euclidean distances between final-layer DistilBERT embeddings and the clean-obfuscated margin $\delta$. Here we note that while the classification performance for DistilBERT exceeds 99\%, the embedding geometry reveals a markedly different picture. We observe that the obfuscated prompts exhibit the largest intra-class variance ($3.33 \pm 6.23$), indicating some embeddings lie extremely close to clean prompts while others are distant. This indicates that there is severe latent-space instability. We also note that inter-class distances to clean prompts have elevated standard deviation ($24.34 \pm 4.51$). This demonstrates partial collapse of obfuscated embeddings onto the clean manifold. Additionally, we observe that the minimal margin $\delta = 1.02$ confirms the presence of \textbf{latent embedding collapse}. $\delta = 1.02$ confirms that some adversarial inputs lie arbitrarily close to clean prompts in embedding space, despite correct predictions. These results quantify the \textbf{performance-robustness gap}, where performance alone would suggest robustness, but the embedding geometry indicates structural fragility.

\begin{figure}[t]
\centering
\includegraphics[width=\columnwidth]{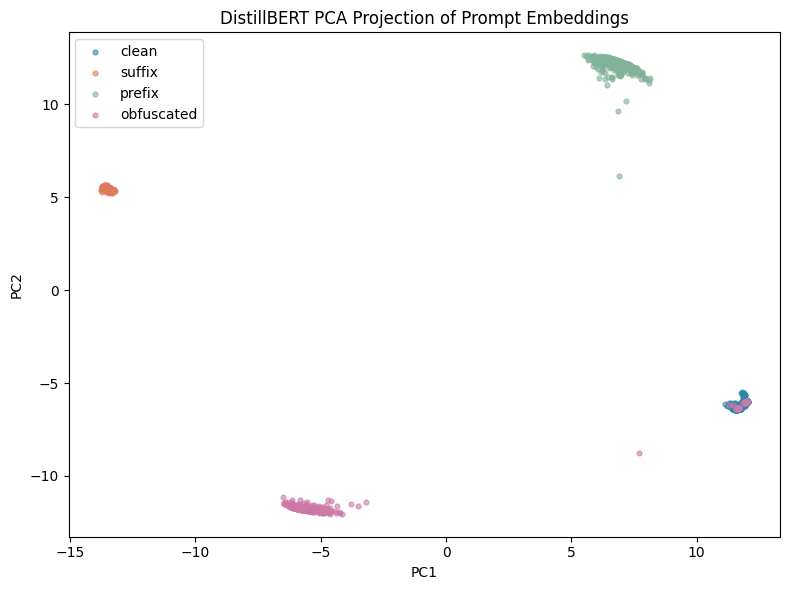}
\caption{DistillBERT PCA Projections of Prompt Embeddings}
\label{fig:distillpca}
\end{figure}

\begin{figure}[t]
\centering
\includegraphics[width=\columnwidth]{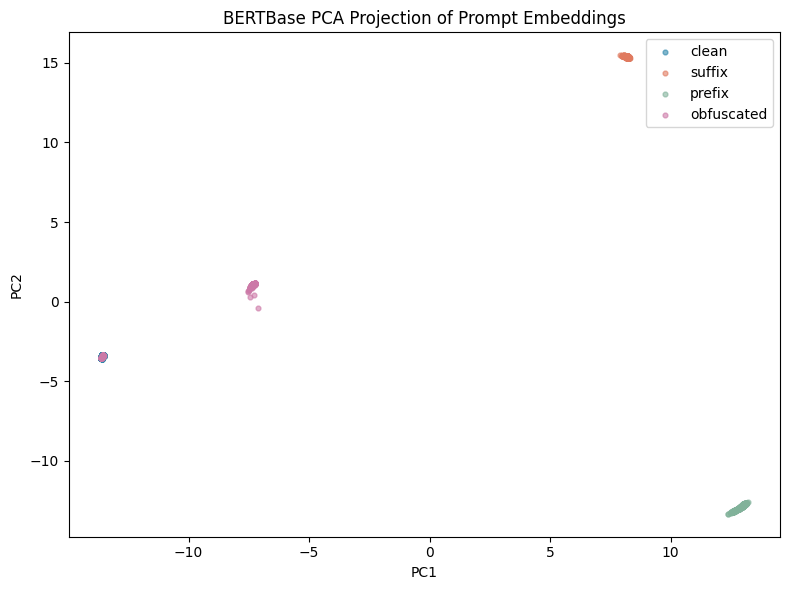}
\caption{BERTBase PCA Projections of Prompt Embeddings}
\label{fig:bertbasepca}
\end{figure}

\begin{figure}[t]
\centering
\includegraphics[width=\columnwidth]{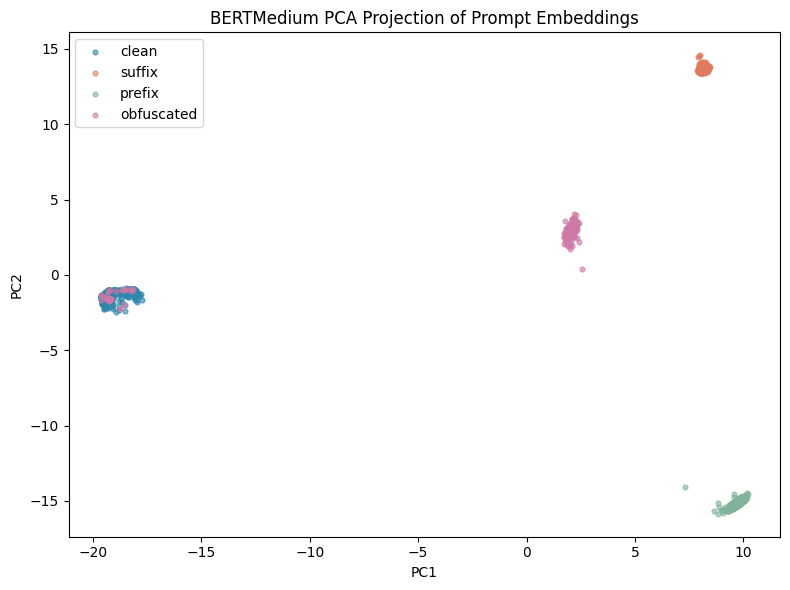}
\caption{BERTMedium PCA Projections of Prompt Embeddings}
\label{fig:bertmediumpca}
\end{figure}

Observing this, we perform a complete evaluation of the 3 models. We analyze the final-layer embeddings of the models through two performance metrics: Principal Component Analysis (PCA) and t-Distributed Stochastic Neighbor Embedding (t-SNE). PCA and t-SNE are chosen for evaluation as they are contemporary metrics used for dimensionality reduction and data visualization in literature \cite{anowar2021conceptual}. For PCA, the results for DistillBERT are in Figure \ref{fig:distillpca}, the results for BERTBase are in Figure \ref{fig:bertbasepca}, and BERTMedium are in Figure \ref{fig:bertmediumpca}. In these results, we note that, despite architectural differences, all the models exhibit the same characteristics. Clean, prefix, and suffix prompts form compact and well-separated manifolds, while obfuscated prompts display elevated dispersion and partial overlap with the clean manifold. Increasing depth or hidden dimensionality does not increase the clean-obfuscated margin nor reduce obfuscated variance.

\begin{figure}[t]
\centering
\includegraphics[width=\columnwidth]{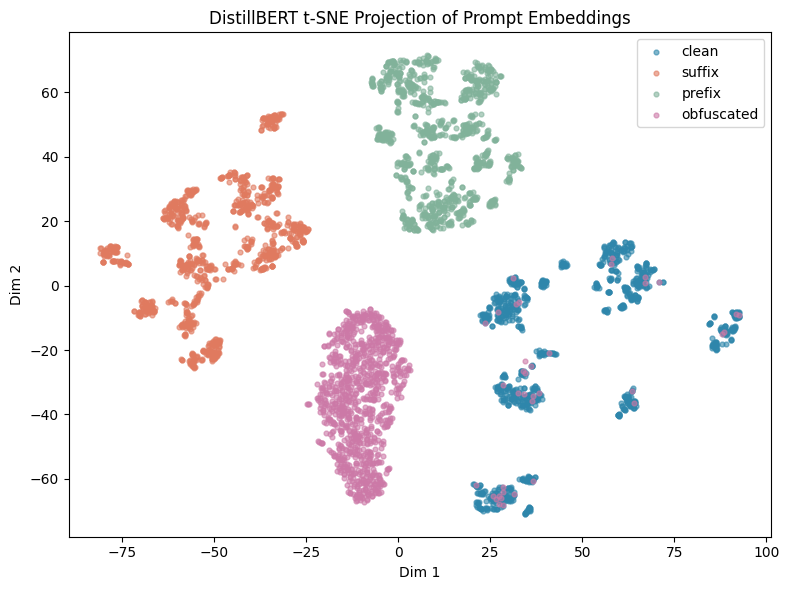}
\caption{DistillBERT t-SNE Projections of Prompt Embeddings}
\label{fig:distilltsne}
\end{figure}

\begin{figure}[t]
\centering
\includegraphics[width=\columnwidth]{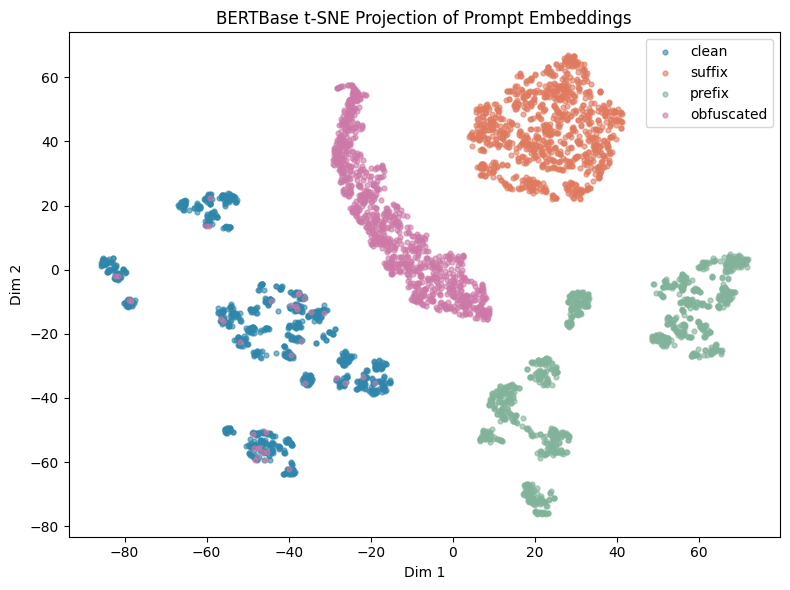}
\caption{BERTBase t-SNE Projections of Prompt Embeddings}
\label{fig:bertbasetsne}
\end{figure}

\begin{figure}[t]
\centering
\includegraphics[width=\columnwidth]{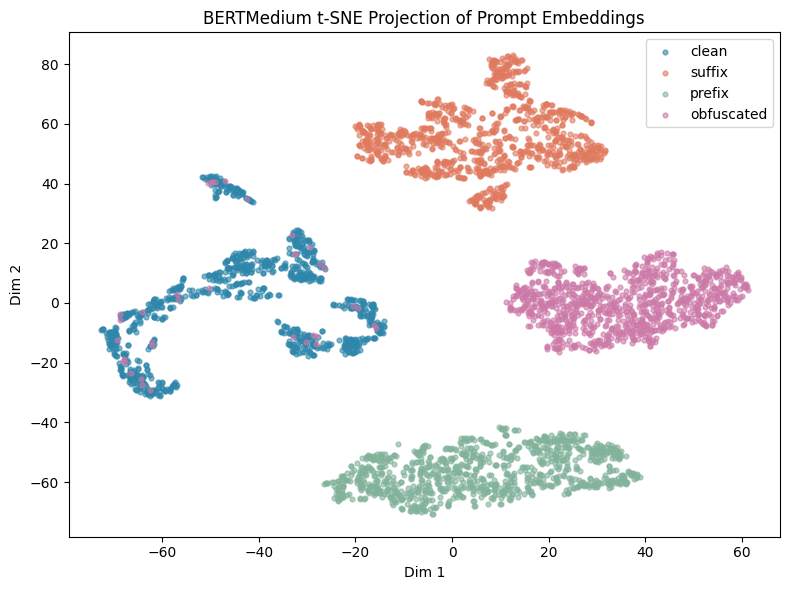}
\caption{BERTMedium t-SNE Projections of Prompt Embeddings}
\label{fig:bertmediumtsne}
\end{figure}

Similarly for t-SNE, the results for DistillBERT are in Figure \ref{fig:distilltsne}, the results for BERTBase are in Figure \ref{fig:bertbasetsne}, and BERTMedium are in Figure \ref{fig:bertmediumtsne}. In these results, we note that, across all models, prefix and suffix prompts remain well separated from clean inputs, which is traditionally consistent with high classification performance. However, obfuscated prompts partially overlap with the clean cluster, providing visual evidence of \textbf{latent embedding collapse} and partial \textbf{manifold collapse}. These geometric intrusions occur even when predictions are correct, reinforcing the premise that performance alone fails to capture latent vulnerability.

These results indicate that architectural scaling improves convergence and classification confidence but does not resolve the underlying geometric fragility. Latent embedding collapse persists across model families, confirming that the \textbf{performance-robustness gap} is not an artifact of insufficient model capacity. Taken together, the quantitative margins, cross-model comparisons, and visualizations demonstrate that high detection performance can coexist with severe geometric fragility. This confirms the central claim of this work: \textbf{robust prompt-injection defense requires geometry-aware evaluation beyond performance-based metrics}.

\section{Conclusion} \label{sec:conclusion}
In this paper, we conducted a geometric analysis of transformer embeddings under multi-operator prompt-injection attacks on LLMs. For this, we formalized the concept of \emph{latent embedding collapse} and connected it to partial manifold collapse in transformer embedding spaces. We also introduced geometric metrics to quantify representational robustness, providing a complementary evaluation to conventional performance metrics. Finally, we demonstrated empirically that increasing model depth or capacity does not resolve latent embedding collapse, indicating that the \textbf{performance-robustness gap} is an inherent phenomenon rather than an artifact of limited model size. Results indicate that multi-operator obfuscated prompts can partially collapse onto the clean embedding manifold, confirming \textbf{latent embedding collapse}, despite near-perfect classification performance. Additionally,  embedding-space metrics, including pairwise distances, intra- and inter-class variance, and the minimal clean-obfuscated margin ($\delta = 1.02$), expose vulnerabilities invisible to standard evaluation, highlighting the \textbf{performance-robustness gap}. Finally, we noted that high classifier performance can coexist with severe structural fragility in the embedding space, emphasizing that true robustness requires geometry-aware evaluation. For future work, we will continue this investigation in multiple directions. One direction is to explore topology-aware descriptors of embedding neighborhoods, since obfuscated prompts may preserve semantic structure while disrupting token-level geometry, causing standard distance metrics to collapse even when topological invariants remain distinct. Another direction will be formalizing the latent robustness gap, which quantifies the discrepancy between classifier confidence and embedding-space stability. Finally, we will design training strategies that explicitly preserve geometric and topological properties under obfuscation.

\bibliographystyle{IEEEtran}
\bibliography{paper}

\end{document}